\documentclass[11pt]{article}
\usepackage{latexsym}

\usepackage{latexsym}
\usepackage{plain}

\usepackage[german,english]{babel}
\usepackage[T1]{fontenc}
\usepackage[latin1]{inputenc}

\usepackage{amsfonts}
\usepackage{amsmath}
\usepackage{mathrsfs}
\usepackage{graphicx}
\usepackage{epstopdf}
	\DeclareGraphicsExtensions{.eps}
\usepackage{multicol}
\usepackage{caption}
\usepackage{amssymb}
\usepackage{array}
\usepackage{rotating}

\numberwithin{equation}{section}

\begin{document}
\def\cl{\centerline}

\cl{\Large{\bf Entropy-power inequality for weighted entropy}}
\vskip 1 truecm
\cl{\Large{\bf Y.~Suhov$^{1-3}$, S.~Yasaei Sekeh$^{4}$, M.~Kelbert$^{5,6}$}}
\vskip .5 truecm

\cl{$^{1}$ DPMMS, University of Cambridge, UK}

\cl{$^{2}$ Math Dept, Penn State University, PA, USA}

\cl{$^{3}$ IPIT RAS, Moscow, RF}

\cl{$^{4}$ DEs, Federal\;University\;of\;S$\tilde{\rm a}$o\;Carlos, SP, Brazil}

\cl{$^{5}$ Dept of Mathematics, University of Swansea, UK}

\cl{$^{6}$ Moscow Higher School of Economics, RF}
\date{\today}

\def\fB{\mathfrak B}\def\fM{\mathfrak M}\def\fX{\mathfrak X}
 \def\cB{\mathcal B}\def\cM{\mathcal M}\def\cX{\mathcal X}
\def\bu{\mathbf u}\def\bv{\mathbf v}\def\bx{\mathbf x}\def\by{\mathbf y}
\def\om{\omega} \def\Om{\Omega}
\def\bbP{\mathbb P} \def\hw{h^{\rm w}} \def\hwi{{h^{\rm w}}}
\def\beq{\begin{eqnarray}} \def\eeq{\end{eqnarray}}
\def\beqq{\begin{eqnarray*}} \def\eeqq{\end{eqnarray*}}
\def\rd{{\rm d}} \def\Dwphi{{D^{\rm w}_\phi}}
\def\BX{\mathbf{X}}\def\Lam{\Lambda}\def\BY{\mathbf{Y}}

\def\mwe{{D^{\rm w}_\phi}}
\def\DwPhi{{D^{\rm w}_\Phi}} \def\iw{i^{\rm w}_{\phi}}
\def\bE{\mathbb{E}}
\def\1{{\mathbf 1}} \def\fB{{\mathfrak B}}  \def\fM{{\mathfrak M}}
\def\diy{\displaystyle} \def\bbE{{\mathbb E}} \def\bu{\mathbf u}
\def\BC{{\mathbf C}} \def\lam{\lambda} \def\bbB{{\mathbb B}}
\def\bbR{{\mathbb R}}\def\bbS{{\mathbb S}} \def\bmu{{\mbox{\boldmath${\mu}$}}}
 \def\bPhi{{\mbox{\boldmath${\Phi}$}}}
 \def\bbZ{{\mathbb Z}} \def\fF{\mathfrak F}\def\bt{\mathbf t}\def\B1{\mathbf 1}

 \def\bp{\bf p}
\def\bq{\bf q}
\def\bR{\bf R}
\def\ovp{\overline{\phi}}
\def\eps{{\epsilon}}

\begin{abstract}
We analyse an analog of the entropy-power inequality for the weighted entropy.
In particular, we discuss connections with weighted Lieb`s splitting inequality and
an Gaussian additive noise formula. Examples and counterexamples are given, for
some classes of probability distributions. \end{abstract}
\vskip .5 truecm



\def\beal{\begin{array}{l}}
\def\beac{\begin{array}{c}}
\def\beacl{\begin{array}{cl}}
\def\ena{\end{array}}
%
%
\section{Introduction. The weighted entropy-power inequality}

Let $x\in\bbR\mapsto\phi (x) \geq 0$ be a given (measurable) function.
The weighted differential entropy (WDE) $\hw_\phi (Z)$ of a real-valued random variable
(RV) $Z$ with a probability density function (PDF) $f_Z$ is defined by the formula
\beq\label{eq:WDE0}\hw_\phi (Z) =\hw_\phi (f_Z):=-\bbE\,\phi (Z)\ln\,f_Z(Z) =-\int_{\bbR^n}\phi (x )f_ Z ( x)\ln\,f_Z(x )\rd x ,\eeq
assuming that the integral is absolutely convergent (with the usual agreement that $0\cdot\ln\,0=0$). Cf.
\cite{BG}, \cite{C}, \cite{SY}. For $\phi (  x )\equiv 1$, the definition yields the standard (Shannon) differential
entropy (SDE). Furthermore, $\phi$ is called a weight function (WF). When we say that $\hw_\phi (Z)$ is finite
we mean that RV $Z$ has a PDF $f_Z$, and the integral in \eqref{eq:WDE0} absolutely converges.



We propose the following bound which we call the weighted entropy-power inequality (WEPI):
for two independent RVs $X_1$ and $X_2$, with $X=X_1+X_2$,
\beq\label{WEPI1}\diy
\exp\left[\frac{2\; \hw_\phi(X_1)}{\bE\,\phi (X_1)}\right]
+\exp\left[\frac{2\; \hw_\phi(X_2)}{\bE\,\phi (X_2)}\right]\leq\exp\left[\frac{2\; \hw_\phi(X)}{\bE\,\phi (X)}\right] ,\eeq
assuming that the WDEs $\hw_\phi( X)$, $\hw_\phi(X_1)$ and $\hw_\phi (X_2)$ are finite, as well as the
expected values $\bE\,\phi (X),\bE\,\phi (X_1),\bE\,\phi (X_2)$ (the latter means that
 $\bE\,\phi (X),\bE\,\phi (X_1),\bE\,\phi (X_2)\in (0,\infty )$).
Again, for $\phi (  x )\equiv 1$, this yields the famous EPI put forward by Shannon; see \cite{CT}, \cite{VG},
 \cite{KS1}, \cite{KS2}. In this note we offer a sufficient condition for \eqref{WEPI1} (see Eqns \eqref{assum1}
and \eqref{eq:WLSI} below); the origins of bound  \eqref{eq:WLSI} go back to Ref. \cite{L}.
We set:
\beq\label{eq:alpha} \alpha=\tan^{-1}\left(\exp\left[\frac{\hw_\phi(X_2)}{\bE\,\phi (X_2)}-\frac{\hw_\phi(X_1)}{\bE\,\phi (X_1)}\right]\right),\;Y_1=\frac{X_1}{\cos\alpha},\;Y_2=\frac{X_2}{\sin\alpha},\eeq
and
\beq\label{eq:zeta}\kappa =\diy\exp\left[\frac{2\hw_\phi(X_1)}{\bE\,\phi (X_1)  }\right]
+\exp\left[\frac{2\hw_\phi(X_2)}{\bE\,\phi (X_2)}\right].\eeq
\vskip .5 truecm\def\phic{\phi_{\rm C}} \def\phis{\phi_{\rm S}}\def\beac{\begin{array}{c}}\def\ena{\end{array}}

{\bf Theorem 1:}\label{lem0:1}
{\sl Given independent RVs $X_1$, $X_2$ and a WF $\phi$,  set $X=X_1+X_2$ and
make the following suppositions. {\rm{(i)}} The expected values obey
\beq \label{assum1}\beac
\diy \bE\,\phi  (X_1)\geq\bE\,\phi (X)\;\hbox{ and }\;
\diy \bE\,\phi (X_2)\geq\bE\,\phi (X)\;\hbox{ if $\kappa\geq 1$,}\\
\diy \bE\,\phi  (X_1)\leq\bE\,\phi (X)\;\hbox{ and }\;
\diy \bE\,\phi (X_2)\leq\bE\,\phi (X)\;\hbox{ if $\kappa\leq 1$.}\ena\eeq

{\rm{(ii)}} With $\alpha$ and $Y_2$, $Y_2$ as in Eqn \eqref{eq:alpha},
\beq\label{eq:WLSI}
 (\cos\alpha)^2 \;h^{\rm w}_{\phic}(Y_1)+(\sin\alpha)^2\; h^{\rm w}_{\phis}(Y_2)\leq
\hw_\phi(X).\eeq
Here
\beq\label{eq:phi12}\phic (x)=\phi(x\cos\alpha),\;\;\phis (x)=\phi(x\sin\alpha)\eeq
and we assume finite WDEs $\hw_\phi(X)$ and
\beqq\hw_{\phic}(Y_2 )=-\bbE\phic (Y_2)\ln f_{Y_2}(Y_2),\;\;
\hw_{\phis}(Y_2 )=-\bbE\phis (Y_2)\ln f_{Y_2}(Y_2).\eeqq
Then WEPI \eqref{WEPI1} holds true.}\\

{\bf{Proof}:} We can write
\beqq \hw_\phi(X_1)=h^{\rm w}_{\phic}(Y_2)+\bE\,\phi (X_1)\log \;\cos\alpha,\;\;
\hw_\phi(X_2)=h^{\rm w}_{\phis}(Y_2)+\bE\,\phi (X_2)\log \;\sin\alpha .\eeqq
Using (\ref{eq:WLSI}), we have the following inequality:
\beqq\beal \hw_\phi(X)\geq (\cos\alpha )^2\Big[\hw_\phi(X_1)-\bE\,\phi (X_1)\log \;\cos\alpha\Big]\\
\qquad\qquad\qquad +(\sin\alpha )^2\Big[\hw_\phi(X_2)-\bE\,\phi (X_2)\log \;\sin\alpha\Big].\ena\eeqq
Furthermore, recalling (\ref{eq:zeta}) we obtain:
\beqq\beal
\hw_\phi(X)\\
\;\;\geq \diy\frac{1}{2\kappa}\Big[\bE\phi (X_1) \log\;\kappa\Big]
\exp\left[\frac{2\hw_\phi(X_1)}{\bE\,\phi (X_1)}\right]
+\frac{1}{2\kappa}\Big[\bE\,\phi (X_2)\log\;\kappa\Big] \exp\left[\frac{2\hw_\phi(X_2)}{\bE\,\phi (X_2)}
\right].\ena\eeqq
By virtue of assumption (\ref{assum1}), we derive:
\beqq \hw_\phi(X)\geq \frac{1}{2} \bE\,\phi (X)\log\;\kappa .\eeqq
The definition of $\kappa$ in Eqn \eqref{eq:zeta} leads directly to the result. $\quad$ $\Box$
\vskip .5 truecm

Paying homage to Ref. \cite{L}, we call the bound \eqref{eq:WLSI} the WLSI (weighted
Lieb$^\prime$s splitting inequality). In the spirit of \cite{L}, the following Theorem 2 can be offered.
(The notation used in Theorem 2 is  self-explanatory; the proof of Theorem 2 is one-line and
omitted.)
\vskip .5 truecm

{\bf Theorem 2:}\;{\sl Let $f$ and $g$ be PDFs on $\bbR$ and $\phi$ a given WF. Assume
that the WDEs\\ $\hw_\phi (f*g)$,  $\hw_\phi (f)$ and $\hw_\phi (g)$ are finite, as well as expected values
$\bbE_f\,\phi$, $\bbE_g\,\phi$. Set
$$\tau=\exp\left[\diy\frac{2\hw_\phi(f)}{\bbE_{f}\,\phi  }\right]+\exp\left[\diy\frac{2\hw_\phi(g)}{\bbE_{g}\,\phi  }
\right].$$
Also suppose that
\beq \beac
\diy \bbE_f\,\phi\geq\bbE_{f*g}\,\phi\;\hbox{ and }\;
\diy \bbE_g\,\phi\geq\bbE_{f*g}\,\phi\;\hbox{ if $\tau\geq 1$,}\\
\diy \bbE_f\,\phi\leq\bbE_{f*g}\,\phi\;\hbox{ and }\;
\diy \bbE_g\,\phi\leq\bbE_{f*g}\,\phi\;\hbox{ if $\tau\leq 1$.}\ena\eeq
and the following inequality holds:
\beq\label{eq 1:3}\begin{array}{l} 2\hw_\phi(f*g)\geq \diy 2\lambda\hw_\phi(f)+2(1-\lambda)\hw_\phi(g)\\
 \qquad\qquad\qquad\qquad- \bbE_f\,\phi  \lambda \log \lambda -\bbE_g\,\phi   (1-\lambda)\log (1-\lambda),\end{array}\eeq
where $\lambda\in [0,1]$ is given by
\beq\diy \lambda=\tau^{-1}\;\exp\left[\diy\frac{2\hw_\phi(f)}{\bbE_{f}\,\phi  }\right].\eeq
Then Eqn \eqref{WEPI1} holds true for independent RVs $X_1$ and $X_2$ where $X_1\sim f$, $X_2\sim g$.}
\vskip .5 truecm

{\bf Remark.} The arguments developed in Section 1 do not use the fact that RVs $X_1$ and $X_2$
possess PDFs. The question of whether the WEPI (as it is presented in Eqn \eqref{WEPI1} or in a modified
form) may hold for cases of discrete distributions
requires a separate investigation. However, constructions used in Section 3 demand existence
of PDFs $f_{X_1}$ and $f_{X_2}$ although some of their technical parts  are valid in a more general
situation.


\section{Examples and counterexamples}
In this section we give several examples where the above inequalities hold or do not hold true.

\vskip .5 truecm
{\bf 2.1. Examples.} First, let us discuss specific conditions equivalent to \eqref{WEPI1}, \eqref{assum1}
or \eqref{eq:WLSI}, for various pairs of RVs. In the next subsection we present results of numerical
simulations showing domains of parameters where Eqns \eqref{WEPI1}, \eqref{assum1} and \eqref{eq:WLSI}
are fulfilled or violated.
\vskip .5 truecm

{\bf 2.1.1.}\;(Normal distributions)\;Let $X_1$, $X_2$ be two independent normal RVs:
$X_1\sim{\rm N}(0, \sigma_1^2)$, $X_2\sim{\rm N}(0,\sigma_2^2)$ and $X\sim{\rm N}(0,\sigma_1^2
+\sigma_2^2)$. Recall (see \cite{SY}, Example 3.1), the WDE $\hw_\phi (Z)$ 
of a normal random variable $Z\sim{\rm N}(0,\sigma^2)$ 
reads
\beqq\hw_\phi (Z)=
\frac{\log \left(2\pi\sigma^2\right)}{2} \bbE\phi (Z) 
+\frac{\log\,e}{2\sigma^2}\bbE Z^2\phi (Z)
;\eeqq
we will use it for $Z=X,X_1,X_2$. The condition $\kappa\geq (\leq )1$ is re-written as
\beq\label{normal.zeta} \diy\sigma_1^2\exp\left\{\frac{\bE [X_1^2\phi (X_1)]}{\sigma_1^2\, \bE\,\phi(X_1)}\right\}+\sigma_2^2\exp\left\{\frac{\bE [X_2^2\phi (X_2)]}{\sigma_2^2\, \bE\,\phi(X_2)}\right\}\geq(\leq)\, (2\pi)^{-1}.\eeq
We have to match it with inequalities
\beqq\bbE\phi (X_1),\;\;\bbE\phi (X_2)\geq (\leq )\bbE\,\phi (X) \eeqq
to fulfill  \eqref{assum1}

To specify the WLSI \eqref{eq:WLSI}, 
we write:
\beq h^{\rm w}_{\phic}(Y_1)=\frac{1}{2}\left[\log\frac{2\pi\sigma_1^2}{(\cos\alpha)^2}\right] \bbE\,\phic (Y_1)+\frac{(\cos\alpha )^2\log e}{2\sigma_1^2}\;\bbE\big[Y_1^2 \phic (Y_1)\big].\eeq
Pluging-in the definition of $\phic$:
\beqq \bbE\phic (Y_1)=\bbE\phi (X_1),\quad \bbE\big[Y_1^2 \phic (Y_1)\big]=\frac{\bbE\big[X_1^2 \phi (X_1)\big]}{(\cos\alpha)^2}.\eeqq
Similar equations hold for $h^{\rm w}_{\phis}(Y_2)$.  Then Eqn \eqref{eq:WLSI} takes the form
\beq\label{eq01.3} \begin{array}{l}
\diy\Big[\log \Big(2\pi (\sigma_1^2+\sigma_2^2)\Big)\Big] \bbE\phi (X)+\frac{\log e}{\sigma_1^2+\sigma_2^2}\; \bbE\big[X^2\phi (X)\big]\\
\quad \geq \diy (\cos\alpha )^2\left[\log\frac{2\pi \sigma_1^2}{(\cos\alpha )^2}\right]\bbE\phi (X_1) +\frac{(\cos\alpha )^2\log e}{\sigma_1^2}\;\bbE [X_1^2 \phi (X_1)]\\
\qquad + \diy (\sin\alpha )^2\left[\log\frac{2\pi \sigma_2^2}{(\sin\alpha )^2}\right]\bbE\phi (X_2)+\frac{(\sin\alpha )^2\log e}{\sigma_2^2}\;\bbE [X_2^2 \phi (X_2)].\end{array}\eeq
\vskip .5 truecm

{\bf 2.1.2.}\;(Gamma-distributions)\;Let $X_1$ and $X_2$ have Gamma distributions, with PDFs $f_{X_i}(x)=\diy\frac{\lambda^{\beta_i}}{\Gamma (\beta_i)} x^{\beta_i-1}e^{-\lambda x}$, $i=1,2$, and $f_X(x)=\diy\frac{\lambda^{\beta}}{\Gamma (\beta )} x^{\beta-1}e^{-\lambda x}$, $x>0$ where $\beta =\beta_1+\beta_2$.
The WDEs are
\beqq\hw_\phi (X_i)=(1-\beta_i)\bbE [\phi (X_i)\log X_i]+\lambda\bbE  [X_i\phi (X_i)]+\log\left(\diy\frac{\Gamma(\beta_i)}{\lambda^{\beta_i}}\right)\bbE\phi (X_i)\eeqq
and similarly for $X$ (with $\beta$ instead of $\beta_i$). The condition $\kappa\geq (\leq )1$ reads
\beq\begin{array}{l} \diy\bigg(\frac{\Gamma{\beta_1}}{\lambda^{\beta_1}}\bigg) \exp\Big\{\frac{\lambda\bE[X_1\phi(X_1)]-(\beta_1-1)\bE[\phi(X_1)\log X_1]}{\bE\,\phi(X_1)}\Big\}\\
\quad+\diy\bigg(\frac{\Gamma{\beta_2}}{\lambda^{\beta_2}}\bigg) \exp\Big\{\frac{\lambda\bE[X_2\phi(X_2)]-(\beta_2-1)\bE[\phi(X_2)\log X_2]}{\bE\,\phi(X_2)}\Big\}\geq (\leq)\,1; \end{array}\eeq
as above, it has to be in conjunction with $\bE\,\phi(X_1),\bE\,\phi(X_2) \geq (\leq)\,\bE\,\phi(X)$.
The WLSI (\ref{eq:WLSI}) takes the following form:
\beq \begin{array}{l}\diy \log \frac{\Gamma(\beta )}{\lambda^\beta}\bE\,\phi(X)-(\beta-1)\bE[\phi(X)\log X]+\lambda\bE[X\phi(X)]\\
\quad \geq \diy \big(\cos\alpha\big)^2\bigg[\lambda\bE[X_1\phi(X_1)]-(\beta_1-1)\bE[\phi(X_1)\log X_1]\bigg]+\bE\,\phi(X_1)\log \frac{\Gamma(\beta_1)}{\lambda^{\beta_1}\cos\alpha}\\
\qquad \diy +\big(\sin\alpha\big)^2\bigg[\lambda\bE[X_2\phi(X_2)]-(\beta_2-1)\bE[\phi(X_2)\log X_2]\bigg]+\bE\,\phi(X_2)\log \frac{\Gamma(\beta_2)}{\lambda^{\beta_2}\sin\alpha}.\end{array}\eeq
\vskip .5 truecm

{\bf 2.1.3.}\;(Exponential distributions)
\;Let $X_1$ and $X_2$ be two independent exponential RVs, with means $\lambda_1^{-1}$ and $\lambda_2^{-1}$,
and WDEs $\hw_\phi (X_i)=\big(\lam_i\log\,\lam_i\big)\bbE\phi (X_i)+\bbE X_i\phi (X_i)$, $i=1,2$. See \cite{SY}, Example 3.2. Then
$f_{X}(x)=\diy\frac{\lambda_1 \lambda_2}{\lambda_1-\lambda_2}\big(e^{-\lambda_2 x}-e^{-\lambda_1 x}\big)$. The inequality $\kappa\geq (\leq )1$ becomes
\beq\label{exp:zeta} \lambda_2^2 \exp\bigg\{\frac{2\lambda_1\bE [X_1\phi (X_1)]}{\bE\, \phi(X_1)}\bigg\}+\lambda_1^2\exp\bigg\{\frac{2\lambda_2\,\bE_{X_2}[X_2\phi]}{\bE\, \phi(X_2)}\bigg\}\geq(\leq)\, \lambda_1^2 \lambda_2^2,\eeq
and to fulfill Eqn \eqref{assum1} it has to be combined with
\beq\label{exp:assum1}\bE\,\phi (X_1),\;\bE\,\phi (X_2)\geq(\leq)\, \diy\frac{\lambda_1 \bE\,\phi (X_2)-\lambda_2 \bE\,\phi(X_1)}{\lambda_1-\lambda_2}.\eeq

In this example, the WLSI \eqref{eq:WLSI}  reads
\beq\begin{array}{l}
\diy \frac{\log \;\lambda_1\lambda_2}{\lambda_1-\lambda_2}\,\Big\{\lambda_2 \bbE\phi (X_1)-\lambda_1 \bbE\phi (X_2)\Big\}+\diy \frac{\lambda_2}{\lambda_1-\lambda_2} \bbE \bigg[\phi (X_1)\log \left(\frac{e^{-\lambda_2 X_1}-e^{-\lambda_1 X_1}}{\lambda_1-\lambda_2}\right)\bigg]\\
\qquad\qquad-\diy\frac{\lambda_1}{\lambda_1-\lambda_2} \bbE\bigg[\phi (X_2) \log \;\left(\frac{e^{-\lambda_2 X_2}-e^{-\lambda_1 X_2}}{\lambda_1-\lambda_2}\right)\bigg]\\
\qquad \geq \diy\lambda_1 (\cos\alpha )^2\,\bbE [X_1\phi (X_1)]-(\cos\alpha )^2\bbE\,\phi(X_1)\log (\lambda_1 \cos\alpha)\\
\qquad \qquad\qquad \qquad\qquad + \diy \lambda_2 (\sin\alpha)^2\bbE [X_2\phi (X_2)]-(\sin\alpha)^2\bbE\,\phi(X_2)\log (\lambda_2 \sin\alpha).\end{array}\eeq
\vskip .5 truecm

{\bf 2.1.4.}\;(Uniform distributions)\;Set $\Phi(x)=\int\limits_0^x\phi(u)\rd u$, $\Phi^*(x)=\int\limits_0^x u\, \phi(u)\rd u$. Let $X_1\sim{\rm U}(a_1,b_1)$ and $X_2\sim{\rm U}(a_2,b_2)$, independently,
where $L_i:=b_i-a_i>0$, $i=1,2$. The WDE $\hw_\phi (X_i)=\diy\frac{\Phi(b_i)-\Phi(a_i)}{L_i}\log \,L_i$.   
Suppose for definiteness that $L_2\geq L_1$ or, equivalently, $C_1:=a_2+b_1\leq a_1+b_2=:C_2$.
Then PDF  $f_{X}$ for $X=X_1+X_2$ has a trapezoidal form with corner points at $A=a_1+a_2$, $C_1$,
$C_2$ and $B=b_1+b_2$:
\beqq f_{X}(x)=\frac{1}{L_1L_2}\times\left\{\begin{array}{cl}0,&\hbox{if $x<A$ or $x>B$,}\\
x-A,&\hbox{if $A<x<C_1$,}\\ L_1,&\hbox{if $C_1<x<C_2$,}\\
B-x,&\hbox{if $C_2<x<B$.}\end{array}\right. \eeqq
The condition $\kappa\geq (\leq)1$ takes the form
\beq\label{eq:Ukappa} L_1^2+L_2^2\geq(\leq)\,1.\eeq

Consider the quantity $\Lam$ (which may be as positive as well as negative):
\beq\begin{array}{l}\diy\Lam= \frac{\log\,L_1}{L_2} \Big[\Phi (C_1)-\Phi (C_2)\Big]\\
\qquad\diy +\frac{1}{L_1L_2}\left[\int_{A}^{C_1}\phi(x)(x-A)\;\log (x-A)\,\rd x
+\diy\int_{C_2}^{B}\phi(x)(B-x)\,\log(B-x)\rd x\right].\end{array}\eeq
Then
\beqq \begin{array}{l} \bbE\,\phi (X)=\diy\frac{1}{L_1L_2}\Bigg\{
\Big[\Phi^*(C_1)-\Phi^*(A)-\Phi^*(B)+\Phi^*(C_2)\Big]\\
\quad-A\Big[\Phi(C_1)-\Phi(A)\Big]+L_1\Big[\Phi (C_2)-\Phi (C_1)\Big]+B\Big[\Phi(B)-\Phi (C_2)\Big]\Bigg\}.
\end{array}.\eeqq
To satisfy condition \eqref{assum1}, we have to assume that
\beq \begin{array}{c}
\diy L_2\big[\Phi(b_1)-\Phi(a_1)\big],\;
\diy L_1\big[\Phi(b_2)-\Phi(a_2)\big]\geq (\leq)\,\bbE\,\phi (X),\end{array}\eeq
in conjunction with bound \eqref{eq:Ukappa}.

The WLSI (\ref{eq:WLSI}) becomes
\beq \begin{array}{l}\hw_\phi (X)=
\diy
\diy -\Lam +\big[\log \big(L_1L_2\big)\big] \bbE\,\phi (X)\\
\qquad \geq \diy (\cos\alpha )^2 \frac{\Phi(b_1)-\Phi(a_1)}{L_1}\log \frac{L_1}{\cos\alpha}
+ \diy (\sin\alpha )^2\frac{\Phi(b_2)-\Phi(a_2)}{L_2}\log \frac{L_2}{\sin\alpha}.\end{array}\eeq

\vskip .5 truecm

{\bf 2.1.5.}\;(A mixed case) Let $X_1$ be a Gamma RV with PDF $f_{X_1}(x)=\diy\frac{\lambda^\beta}{\Gamma(\beta)} x^{\beta-1}e^{-\lambda x}$ and the cumulative distribution function $F_{X_1}(x)$. The
WDE $\hw_\phi (X_1)$ has been specified in Example 2.1.3. Take RV $X_2$ from the uniform distribution
${\rm U}(a,b)$, where $L:=b-a>0$, independent of $X_1$. The WDE  $\hw_\phi (X_1)$ has been specified in
Example 2.1.4. We can write
\beqq \label{Conv:Gamma.Uni} f_{X}(t)=\diy \frac{1}{L}\Big[F_{X_1}(t-a)
-F_{X_1}(t-b)\Big].\eeqq
As in Example 2.1.3, let $\Phi(x)=\int_0^x\phi (u)\rd u$. Next, set:
\beq\begin{array}{c}
\Theta =\diy \int_0^\infty \phi(x) \left[F_{X_1}(x-a)-F_{X_1}(x-b)\right]\log\Big[F_{X_1}(x-a)-F_{X_1}(x-b)\Big]\
\rd x.\end{array}\eeq
Then
\beq\label{WEconv} \begin{array}{l}h^{\rm w}_\phi(X)
=\diy\frac{\log(L\,\Gamma(\beta ))}{L}\\
\diy\qquad\times\bbE\Big[\Phi(X_1+b){\mathbf 1}(X_1>-b)-\Phi(X_1+a){\mathbf 1}(X_1>-a)\Big]-
\diy\frac{\Theta}{L}.\end{array}\eeq
The quantity $\kappa$ is specified by
\beq\label{mix.zeta}\diy \kappa =\left(\frac{\Gamma(\beta)}{\lambda^\beta}\right)^2
\exp\Big[\frac{\lambda\bE [X_1\phi (X_1)]-(\beta-1)\bE [\phi (X_1)\log X_1]}{\bE\,\phi(X_1)}\Big]
+L^2. \eeq
Note that if $b-a\geq 1$ we always have $\kappa >1$. To fulfill condition  \eqref{assum1},
we have to assume that
\beq \left\{\begin{array}{c}\diy L\bbE\,\phi(X_1)\geq(\leq)\, \bbE [\Phi (X_1+b)-\Phi (X_1+a)]\\
\diy\Phi (b)-\Phi (a)\geq(\leq)\, \bbE [\Phi (X_1+b)-\Phi (X_1+a)]\;
 \end{array}\right.\hbox{depending on $\kappa\geq (\leq)1$.}\eeq

The WLSI \eqref{eq:WLSI} takes the form:
\beq \begin{array}{l}
\hw_\phi (X):=\log[L\Gamma(\beta)]\,\bbE\Big[\Phi(X_1+b){\mathbf 1}(X_1>-b)
-\Phi (X_1+a){\mathbf 1}(X_1>-a)\Big]-\diy\Theta \\
\quad\geq L (\cos\alpha)^2 \left[(1-\beta)\bbE [\phi (X_1)\log X_1]+\lambda\bbE  [X_1\phi (X_1)]+\log\left(\diy\frac{\Gamma(\beta)}{\lambda^{\beta} \cos\alpha}\right)\bbE\phi (X_1)\right]\\
\qquad\qquad + \diy (\sin\alpha)^2 \frac{\Phi(b)-\Phi(a)}{b-a}\log \left(\frac{L}{\sin\alpha}\right).\end{array}\eeq

{\bf 2.1.6.}\;(Cauchy distributions)\;Let $X_1$, $X_2$ be independent, with PDFs $f_{X_j}(x)$ $=$ $(\pi\theta_j)^{-1}$\\ $\times\Big[1+(x-\mu_j)^2/\theta_j^2\Big]^{-1}$, $x\in\bbR$, $j=1,2$. Then $f_{X_j}(x)$ is
f the same form,
with parameters $\mu=\mu_1+\mu_2$ and $\theta =\theta_1+\theta_2$. For the WDEs $\hw_\phi(X)$
we have the formula
\beqq\hw_\phi(X)=\bbE\,\phi(X)\log(\pi\theta) +\bbE\left[\phi (X)\log (1+(X-\mu )^2/\theta^2)\right]
\eeqq
and similarly for $\hw_\phi(X_i)$.
The condition $\kappa\geq (\leq )1$ is re-written as
\beq\label{cauch.zeta}\begin{array}{l}
\diy\theta_1^2\exp\Big\{\frac{2 \bE[\phi (X_1)\log(1+(X_1-\mu_1)^2/\theta_1^2)]}{\bE\,\phi(X_1)}\Big\}\\
\qquad+\diy \theta_2^2\,\exp\Big\{\frac{2 \bE [\phi
(X_2)\log(1+(X_2-\mu_2)^2/\theta_2^2)]}{\bE\,\phi(X_2)}\Big\} \geq(\leq)\, \pi^{-2};\end{array}\eeq
to satisfy \eqref{eq:WLSI} we have to match it with $\bbE\phi (X_1),\;\bbE\phi (X_2)\geq (\leq )\bbE\,\phi (X)$.
The WLSI \eqref{eq:WLSI} reads
\beq \label{eq:cauchy}\begin{array}{l}\diy\bbE\,\phi(X)\log(\pi\theta) +\bbE\left[\phi (X)\log (1+(X-\mu )^2/\theta^2)\right]\\
\quad\geq \diy (\cos\alpha )^2 \bbE\left[\phi (X_1)\log (1+(X_1-\mu_1)^2/\theta_1^2\right]+
(\cos\alpha )^2 \bbE\,\phi(X_1)\log \frac{\pi \theta_1}{\cos\alpha}\\
\qquad\quad +\diy(\sin\alpha )^2 \bbE\left[\phi (X_2)\log (1+(X_2-\mu_2)^2/\theta_2^2)\right]+(\sin\alpha ^2)\bbE\,\phi(X_2)
\log \frac{\pi\theta_2}{\sin\alpha}.\end{array}\eeq
\vskip .5 truecm

{\bf 2.1.7.}\;(A distribution with an infinite SDE) Here we take independent $X_1,X_2\sim g$ where $g$ is a PDF
on $(1,\infty )$:
$$g(x)=\frac{1}{x\big[(\ln\,x)^2+1\big]},\;\;x>1.$$
Here we have that
\beqq f_{X}(t)=\int g(t-s)g(s) \rd s =\int_1^{t-1} \frac{1}{(t-s)\bigg[1+(\ln(t-s))^2\bigg]s\bigg[
1+(\ln s)^2\bigg]}\rd s. \eeqq
Therefore
\beq\label{eq:hwphiX}\begin{array}{l} h^{\rm w}_\phi(X)\\
=\diy -\int_2^\infty \phi(x)\int_1^{x-1} \frac{1}{(x-s)\bigg[1+(\ln(x-s))^2\bigg]s\bigg[1+(\ln s)^2\bigg]}\rd s\\
\qquad \diy\times \log \Bigg(\int_1^{x-1} \frac{1}{(x-w)\bigg[1+(\ln(x-w))^2\bigg]w\bigg[1+(\ln w)^2\bigg]}
\rd w\Bigg)\rd x,\end{array}\eeq
assuming that WF $\phi$ decreases fast enough so that the integral in \eqref{eq:hwphiX}
absoltely converges.
The bound $\kappa \geq(\leq )\,1$ now reads:
\beq\label{g.zeta} \diy \exp\left\{\frac{2\bE \Big[\phi (X_1)\log\big\{X_1[1+(\ln X_1)^2+1]\big\}\Big]}{
\bE\,\phi(X_1)}\right\}\geq(\leq)\, \frac{1}{2}\,,\eeq
and we again have to match it with inequality
$\bbE\phi (X_1),\;\;\bbE\phi (X_2)\geq (\leq )\bbE\,\phi (X)$.

The WLSI \eqref{eq:WLSI} takes the form:
\beq \begin{array}{l}
\diy-\int_2^\infty \phi(x)\int_1^{x-1} \frac{1}{(x-s)\bigg[1+(\ln(x-s))^2\bigg]s\bigg[1+(\ln s)^2\bigg]}\rd s\\
\quad \diy \times\log \Bigg(\int_1^{x-1} \frac{1}{(x-w)\bigg[1+(\ln(x-w))^2\bigg]w\bigg[1+(\ln w)^2\bigg]}\rd w\Bigg)\rd x\\
\\
\qquad\qquad\geq \bbE \big[\phi (X_1)\log {X_1}\big]+\bbE\Big\{\phi (X_1)\log\big[1+(\ln X_1)^2)\big]\Big\}\\
\qquad\qquad\qquad\qquad -\bbE\,\phi (X_1) \Big[(\cos\alpha )^2\log \cos\alpha+(\sin\alpha )^2\log \sin \alpha\Big].\end{array}\eeq
\vskip .5 truecm

{\bf 2.2. Numerical results.} As was said, in this subsection we comment on some numerical evidence that
(i) Eqn \eqref{WEPI1} does not always hold true, and (ii)  assumptions \eqref{assum1} \eqref{eq:WLSI} in Theorem 1
are not necessary for the WEPI \eqref{WEPI1}.
Our observations are of a preliminary character, and we think that further numerical simulations are needed
here, to build a detailed picture.

\vskip .5 truecm

{\bf 2.2.1.}\;(Normal distributions)\;Assume that $X_1$ and $X_2$ are normal RVs as in
Example 2.1.1. Choose $\phi (x)=|x^2-2|$. The graph in Figure 2.2.1 presetns the difference between
the RHS and the LHS in \eqref{WEPI1}: when this difference is non-negative, the WEPI is satisfied,
otherwise the WEPI fails.The graph shows that there is a domain of parameter
$(\sigma_1,\sigma_2)$ where Eqn \eqref{WEPI1} does not hold. Additional
simulations state that there is a domain where
\eqref{WEPI1} holds and only one of Eqns \eqref{assum1}, \eqref {eq:WLSI} is satisfied. For example,
in the square where $0.55 <\sigma_1,\sigma_2<0.551$ the condition \eqref {eq:WLSI} is violated but \eqref{assum1}, \eqref{WEPI1} hold true.

\vskip .5 truecm

{\bf 2.2.2.}\;(Gamma distributions)\;Assume that $X_1$ and $X_2$ are gamma RVs as in
Example 2.1.2. Here we choose $\phi (x)=x e^{-x}$. The graph in Figure 2.2.2
again shows the difference between the RHS and the LHS in \eqref{WEPI1}: Here the WEPI is
satisfied (in the presented range of parameters $(\beta_1,\beta_2)$). As in Example 2.2.1,
additional simulations assert that there is a domain where
\eqref{WEPI1} holds while none of  \eqref{assum1}, \eqref {eq:WLSI} is satisfied. For example,
in the square $0.01<\beta_1<0.1$ and $5<\beta_2<6.1$ both conditions  \eqref{assum1}, \eqref {eq:WLSI} are violated but  \eqref{WEPI1}
holds true.
\begin{figure}[!htbp]
\begin{center}
\includegraphics[scale=1]{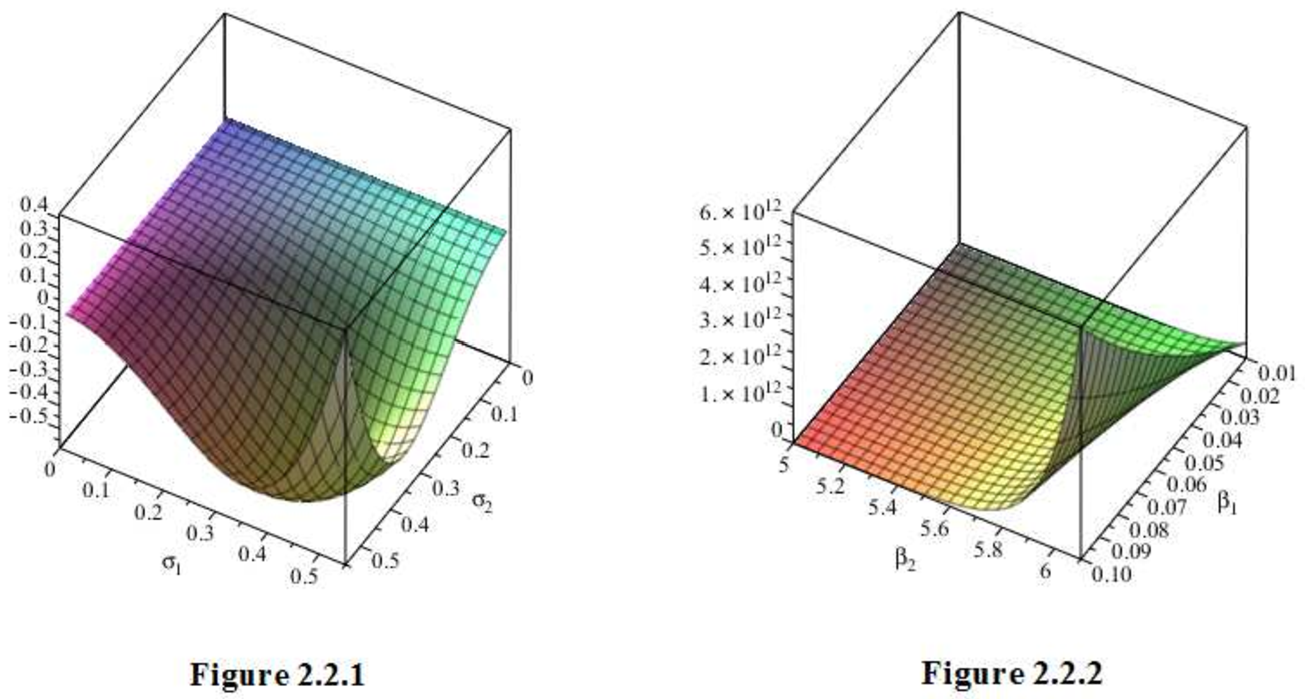}
\end{center}
\label{fig}
\end{figure}

\vskip .5 truecm

\vskip .5 truecm

\section{WDE and an additive Gaussian noise}

{\bf 3.1. Integral representations for WDEs.} Following \cite{VG}, \cite{KS1}, the WLSI \eqref{eq:WLSI}
can be re-written (under certain
conditions on $f_{X_2}$, $f_{X_2}$ and $\phi$) in terms of integral
representations of the entropies $\hw_\phi(X)$,
$h^{\rm w}_{\phic}(Y_1)$ and $h^{\rm w}_{\phis}(Y_2)$: cf. Eqns  \eqref{eq:formula1},
\eqref{eq:formula2} below. 
Despite its cumbersome appearance, formulas \eqref{eq:formula1} and \eqref{eq:formula2} have
an advantage: they does not include logarithms. (However, note condition \eqref{eq:cond3}.) Throughout the presentation in this section, the reader can notice persistent similarities with \cite{KS1}.

In this section we work with a two-variable WF $(x,y)\in\bbR\times\bbR\mapsto \rho (x,y)\geq 0$ and
a number of reduced WDEs involving various integrals of $\rho$.
Let $Z$ and $N$ be two independent RVs, where $N\sim{\rm{N}}(0,1)$ with standard normal PDF $f^{\rm{No}}$, while $Z$
has a PDF $f_Z$. Following \cite{VG} and \cite{KS1},  RV $Z$ will represent a signal and $N$ an (additive)
Gaussian noise; RV $Z$ will be a pre-cursor for $X=X_1+X_2$, $Y_1$ and $Y_2$. Given $\gamma>0$, $y\in\bbR$, set:
\beq\label{eq:xietabeta}\begin{array}{c} \xi_Z(y,\gamma)=\diy\frac{\diy\int (y-t\sqrt{\gamma})y
f^{\rm{No}}(y-t\sqrt{\gamma})f_Z(t)\rd t}{\diy\int f^{\rm{No}}(y-z\sqrt{\gamma})f_Z(z)\rd z},\\
\eta_Z(y,\gamma)=\diy\frac{\diy\int (y-w\sqrt{\gamma})w f^{\rm{No}}(y-w\sqrt{\gamma})f_Z(w)\rd w}{
\diy\int f^{\rm{No}}(y-z\sqrt{\gamma})f_Z(z)\rd z},\\
\zeta_Z(x,y,\gamma )=\diy\int_{-\infty}^y \rho(x,v)(v-x\sqrt{\gamma}) x \frac{f_{Z,Z\sqrt{\gamma}+N}(x,v)}{
f_{Z,Z\sqrt{\gamma}+N}(x,y)} \rd v.\end{array}\eeq
 \vskip .5 truecm\def\beac{\begin{array}{c}}

{\bf Theorem 3.} {\sl Let $X_1$, $X_2$, $N$ and $N^\prime$  be independent RVs, with $X=X_1+X_2$, where {\rm{(i)}}
$X_j$ are with bounded and continuous PDFs $f_{X_j}$ such that $\bbE |\log f_{X}(X)|<\infty$ and $\bbE |\log f_{X_j}(X_j)|<\infty$, $j=1,2$, and {\rm{(ii)}} $N,N^\prime\sim{\rm N}(0,1)$.
Assume that for $Z=X_1$, $Z=X_2$ and $Z=X$,  the conditional expectation
$\bbE\left[f_Z\left(Z+\diy\frac{N-N^\prime}{\sqrt\gamma}\right)\Big|Z,N\right]$ is such that, for some
integrable RV $\chi (Z,N)\geq 0$
\beq\label{eq:cond3}\left|\,\log\bbE\left[f_Z\left(Z+\frac{N-N^\prime}{\sqrt\gamma}\right)\Big|Z,N\right]
\,\right|\leq \chi (Z,N).\eeq

Next, consider a WF $(x,y)\in\bbR\times\bbR\mapsto \rho (x,y)$. Suppose that $\rho$
is continuous and bounded, and $\forall$ $x\in\bbR$, $\exists$ a limit $\phi (x)=\lim\limits_{y\to\pm\infty}\rho(x,y)$.
Introduce additional WFs\def\rhoc{\rho_{\rm C}} \def\rhos{\rho_{\rm S}}
\beq\beac\rhoc (x,y)=\rho (x\cos\vartheta,y),\;\rhos (x,y)=\rho (x\sin\vartheta,y),\hbox{
with $\phi_{{\rm C}/{\rm S}}(x)=\lim\limits_{y\rightarrow\pm\infty}\rho_{{\rm C}/{\rm S}}(x,y)$,}\\
\phi^*(v)=\int \rho(x,v) f_{X}(x)\rd x,\;
\diy\phi^*_{1}(v)=\int \rhoc
(x,v) f_{Y_1}(x)\rd x,\;\phi^*_{2}(v)=\int \rhos (x,v) f_{Y_2}(x)\rd x\ena\eeq
where $Y_2$, $Y_2$ are as in \eqref{eq:alpha}.

Then, the WDE $\hw_\phi (X)$ of the sum $X=X_1+X_2$ in the RHS of \eqref{eq:WLSI} admits the representation
\beq\label{eq:formula1}\beal
\diy \int_0^\infty \diy \frac{1}{2\sqrt{\gamma}}\;\bbE\Big[\zeta_{X}
\Big(X, X
\sqrt{\gamma}+N,\gamma\Big)\diy\xi_{X}\Big(X\sqrt{\gamma}+N,\gamma\Big)\\
\qquad\qquad\qquad -\diy\rho\Big(X, X\sqrt{\gamma}+N\Big)
\diy\eta_{X}\Big(X\sqrt{\gamma}+N,\gamma\Big)\Big] \rd \gamma+\diy h^{\rm w}_{\phi^*}(N).\ena
\eeq
On the other hand, for $h^{\rm w}_{\phic}(Y_1)$ and $h^{\rm w}_{\phis}(Y_2)$ we have the respective formulas
\beq\label{eq:formula2}\beal
\diy \int_0^\infty \diy \frac{1}{2\sqrt{\gamma}}\;\bbE\Big[\zeta_{Y_j}\Big(Y_j, Y_j
\sqrt{\gamma}+N,\gamma\Big)\diy\xi_{Y_j}\Big(Y_j\sqrt{\gamma}+N,\gamma\Big)\\
\qquad\qquad -\diy\rho\Big(Y_j, Y_j\sqrt{\gamma}+N\Big)
\diy\eta_{Y_j}\Big(Y_j\sqrt{\gamma}+N,\gamma\Big)\Big] \rd \gamma+\diy h^{\rm w}_{\phi^*_1}(N),\;j=1,2.\ena
\eeq}
 \vskip .5 truecm
The proof of Theorem 3 uses two technical assertions,
Lemmas 3.1 and 3.2. They address the cases $\gamma =0$ and $\gamma \to\infty$ (that is, the
integration endpoints in  \eqref{eq:formula1} and \eqref{eq:formula2}). For the definition of the weighted
conditional and mutual entropies, see Eqns (1.11), (1.12) in \cite{SY}.
 \vskip .5 truecm

{\bf Lemma 3.1.} (Cf. Lemma 2.4 in \cite{KS1}.) {\sl Let $Z$, $U$ be independent RVs. Assume that $U$ has a
bounded and continous PDF $f_U\in{\rm C}^0(\bbR^d)$: $\int f_U(  x)\rd x =1$ and
${\rm{ess}}\,\sup[f_{U}(  x),  x\in \bbR^d]<+\infty$. The distribution of $Z$ may have discrete and continous parts;
we refer to the PMF $f_Z(x)$ relative to a reference measure $\nu (\rd x)$. Next, suppose that a bounded WF
$\;(x,y)\in\bbR\times\bbR\mapsto \rho (x,y)$ has been given and assume that
$\bbE|\log f_Z(Z)|<+\infty$.
Consider the weighted mutual entropy (WME)\\ $i^{\rm w}_{\rho}(Z:\sqrt{\gamma}Z+U)$ between $Z$
and $\sqrt{\gamma}Z+U$ where $\gamma >0$ is a parameter. Then}
\beq\label{eq:limiphi}\lim\limits_{\gamma\rightarrow 0}i^{\rm w}_{\rho}(Z:\sqrt{\gamma}Z+U)=0.\eeq
\def\orho{{\overline\rho}}

{\bf Proof.}\;According to the definition of the WME, for a pair of RVs $Z,V$ with a conditional
PDF $f_{V|Z}(y,x)$ we have an equality involving a weighted conditional entropy (WCE)
\beqq i^{\rm w}_{\rho}(Z:V)=h^{\rm w}_{\psi_Z}(Z)-\hw_\rho(Z|V)\;\hbox{where}\;\psi_Z(x)=\int \rho(x,y)
f_{V|Z}(y|x)\rd y.\eeqq
Setting $V=\sqrt{\gamma}Z+U$, we can write a representation for the WCE:
\beqq \begin{array}{l}\hw_\rho(Z|\sqrt{\gamma}Z+U)=\diy\int \rho(x,y) f_U(y-\sqrt{\gamma}x)f_Z(x)\\
\qquad\qquad\qquad\qquad\qquad\times\diy\log\left[\frac{\int f_U(y-\sqrt{\gamma}z)f_Z(z) dz}{f_U(y-\sqrt{\gamma}x)f_Z(x)}\right] dy \nu(\rd x).\end{array}\eeqq
Using that $\rho$ and $f_U$ are bounded, with the help of the Lebesgue dominated convergence theorem
we have that as $\gamma\rightarrow 0$, the ratio under the log converges to $[f_Z(x)]^{-1}$. Consequently,
\beqq \lim\limits_{\gamma\rightarrow 0}\hw_\rho(Z|\sqrt{\gamma}Z+U)=h^{\rm w}_{\psi^*_Z}(Z)\;
\hbox{where}\;\psi^*_Z(x)=\int\rho(x,y)f_U(y)\rd y.\eeqq
Moreover, with $\psi^*_{Z,\gamma}(x)=\diy\int \rho(x,y)f_U(y-\sqrt{\gamma}x) \rd y$, we introduce:
\beqq h^{\rm w}_{\psi^*_{Z,\gamma}}(Z)=-\int \rho(x,y) f_U(y-\sqrt{\gamma}x) f_Z(x)\log f_Z(x) \rd y \nu(\rd x).\eeqq
At this stage we again apply the Lebesgue dominated convergence theorem and deduce that
$\lim\limits_{\gamma\rightarrow 0}h^{\rm w}_{\psi^*_{Z,\gamma}}(Z)=h^{\rm w}_{\psi^*_Z}(Z)$. This leads
to \eqref{eq:limiphi}. $\qquad$ $\Box$
\vskip .5 truecm

{\bf Lemma 3.2.}  (Cf. Lemma 4.1 in \cite{KS1}.)\;{\sl Let $Z$, $U$ and $U^\prime$ be independent RVs, $Z$ with a PDF $f_Z$ and
$U,U^\prime$ with a PDF $f_U$.
As before, consider a WF $(x,y)\in\bbR\times\bbR\mapsto \rho (x,y)$. Suppose that $f_Z$ and $\rho$
are continuous and bounded, and there exists a limit $\orho (x)=\lim\limits_{y\to\infty}\rho(x,y)$.
Next, assume that for some RV $\chi (Z,U)\geq 0$ with $\bbE \chi (Z,U)<\infty$, we have
$\left|\,\log\bbE\left[f_Z\left(Z+\frac{U-U^\prime}{\sqrt\gamma}\right)\Big|Z,U\right]
\,\right|\leq \chi (Z,U)$.
Then}
\beq \label{eq:Lem3.2}\begin{array}{l}\diy\hw_\orho (Z)=\lim\limits_{\gamma\rightarrow\infty}\big[i^{\rm w}_{\rho}(Z:\sqrt{\gamma}Z+U)+h^{\rm w}_{\psi^*_{U,\gamma}}(U)\big]\;\hbox{\sl where}\\
\quad\diy\hbox{$h^{\rm w}_{\psi^*_{U,\gamma}}(U)=-\bbE\psi^*_{U,\gamma}(U)\log\,f_U(U)$ and
 $\psi^*_{U,\gamma }(u)=\diy\int \rho(x,u+\sqrt{\gamma}x)f_Z(x) dx$.}\end{array}\eeq
\def\rw{{\rm w}}

{\bf Proof.}\;We can write 
\beq\label{eq:Lem3.2p}\begin{array}{l} i^\rw_\rho (Z:\sqrt{\gamma}Z+U)+h^\rw_{\psi^*_{U,\gamma}}(U)\\
\qquad\diy =-\diy\int \rho(x,u+\sqrt{\gamma}x) f_U(u) f_Z(x)\log \left[\int f_U(v)f_Z\left(x+\frac{u-v}
{\sqrt{\gamma}}\right)\rd v\right]\rd u\rd x.\end{array}\eeq
Passing to the limit $\gamma \rightarrow \infty$, Eqn \eqref{eq:Lem3.2p} yields
\eqref{eq:Lem3.2}, again owing to the Lebesgue
dominated convergence theorem. $\qquad$ $\Box$
\vskip .5 truecm

{\bf Proof of Theorem 3.}\; We again use $Z$ as a substitute for RVs $Y_1$, $Y_2$ and $X=X_1+X_2$ .
Given $\gamma>0$, write the joint WDE for $Z$ and $Z\sqrt{\gamma}+N$:
\beq\label{join:WE2}\begin{array}{l}\hw_\rho(Z,Z\sqrt{\gamma}+N)
=-\diy\int \rho(x,x\sqrt{\gamma}+v)f^{\rm{No}}(v)f_Z(x)\log f^{\rm{No}}(v) \rd x\rd v\\
\qquad-\diy\int \rho(x,x\sqrt{\gamma}+v)f^{\rm{No}}(v)f_Z(x)\log f_Z(x) \rd x\rd v
=h^{\rm w}_{\psi^{(1)}_{Z,\gamma}}(Z)+h^{\rm w}_{\psi^{(2)}_{N,\gamma}}(N).\end{array}\eeq
Here and below, $\forall$ $\gamma ,\theta >0$,
\beq \label{psi12} \psi^{(1)}_{Z,\gamma}(x)=\diy\int \rho(x,x\sqrt{\gamma}+v) f^{\rm{No}}(v) \rd v,\;\;
\psi^{(2)}_{N,\theta}(v)=\diy\int \rho(x,x\sqrt{\theta}+v) f_Z(x) \rd x,\eeq
with $h^{\rm w}_{\psi^{(2)}_{N,\theta}}(N)=-\int\left[\int \rho(x,x\sqrt{\theta}+v) f_Z(x)\rd x\right]f^{\rm{No}}(v)
\log\,f^{\rm{No}}(v)\rd v$.

Moreover, according to Lemma 3.2 (with $U=N$, $U^\prime =N^\prime$), $\forall$ $\eps >0$ we have
\beqq\begin{array}{cl}\diy \hw_\orho(Z)=\diy\int_\eps^\infty \frac{\rd}{\rd\gamma}\big[i^{\rm w}_\rho(Z:Z\sqrt{\gamma}+N)+h^{\rm w}_{\psi^{(2)}_{N,\gamma}}(N)\big]\rd \gamma+\diy i^{\rm w}_\rho(Z:Z\sqrt{\eps}+N)
 +h^{\rm w}_{\psi^{(2)}_{N,\eps}}(N).\end{array}\eeqq
To analyze the WDE $h^{\rm w}_{\psi^{(2)}_{N,\gamma}}(N)$, we use Lebesgue$^\prime$s dominated convergence theorem.  This yields
\beqq \lim\limits_{\eps\rightarrow 0}h^{\rm w}_{\psi^{(2)}_{N,\eps}}(N)=h^{\rm w}_{\rho^*_N}(N)
\;\hbox{ where }\;\rho^*_N(v)=\diy\int \rho(x,v) f_Z(x)\rd x.\eeqq
In addition we get that the WME $ i^{\rm w}_\rho(Z:Z\sqrt{\gamma}+N)$ is represented as the difference
\beq h^{\rm w}_\psi(Z\sqrt{\gamma}+N)-h^{\rm w}_{\psi^{(2)}_{N,\gamma}}(N)
\hbox{ with }\psi(x)=\int \rho(x,v)f^{\rm{No}}(v-\sqrt{\gamma} x)\rd v.\eeq
Note that $\psi^{(2)}_{N,\gamma}=\psi^*_{U,\gamma}$; see (\ref{eq:Lem3.2}). Furthermore, owing to
Lemma 3.1 (with $U=N$), we write:
\beq\label{eq:Equival.WE} \hw_\orho(Z)=\diy \int_0^\infty \frac{\rd}{\rd\gamma} \hw_\psi(Z\sqrt{\gamma}+N) \rd \gamma+h^{\rm w}_{\rho^*_N}(N),\;\hbox{still with $\orho (x)=\lim\limits_{y\to\infty}\rho(x,y)$.}\eeq

We are now going to analyze the derivative $\diy\frac{\rd}{\rd\gamma} h^{\rm w}_\psi(Z\sqrt{\gamma}+N)$ representing it as
\beq\label{eq:07} \begin{array}{l}\diy
-\diy\frac{\rd}{\rd\gamma}\int\int \rho(x,y)f^{\rm{No}}(y-x\sqrt{\gamma})\; f_Z(x)
\log\left[\int f^{\rm{No}}(y-t\sqrt{\gamma})\; f_Z(t)\rd t\right] \rd x\rd y\\
\quad =-\diy\frac{1}{2\sqrt{\gamma}}\int\int \rho(x,y) f_Z(x) f^{\rm{No}}(y-x\sqrt{\gamma})(y-x\sqrt{\gamma})x\\
\qquad\qquad\diy\times\log \left[\int f^{\rm{No}}(y-t\sqrt{\gamma})\; f_Z(t)\rd t\right] \rd x\rd y\\
\qquad -\diy\frac{1}{2\sqrt{\gamma}}\int\int\rho(x,y) f_Z(x) f^{\rm{No}}(y-x\sqrt{\gamma})\\
\qquad\qquad\times
\diy\bigg[\frac{\int w(y-w\sqrt{\gamma}) f^{\rm{No}}(y-w\sqrt{\gamma})f_Z(w)\rd w}{\int f^{\rm{No}}(y-z\sqrt{\gamma})f_Z(z)\rd z}\bigg]\rd x\rd y.\end{array}\eeq
The first integral in the RHS of \eqref{eq:07} is done by parts. This leads to the following expression:
\beqq\label{eq:08} \begin{array}{l}
\diy\frac{1}{2\sqrt{\gamma}}\int\int_{-\infty}^y\int\rho(x,v) f_Z(x) f^{\rm{No}}(v-x\sqrt{\gamma})
(v-x\sqrt{\gamma})x\rd x\rd v\\
\qquad\qquad\qquad\times\diy\bigg[\frac{\int(y-t\sqrt{\gamma})y f^{\rm{No}}(y-t\sqrt{\gamma})f_Z(t)\rd t}{\int f^{\rm{No}}(y-z\sqrt{\gamma})f_Z(z)\rd z}\bigg]\rd y\\
\quad-\diy\frac{1}{2\sqrt{\gamma}}\int\int\rho(x,y) f_Z(x) f^{\rm{No}}(y-x\sqrt{\gamma})
\diy\bigg[\frac{\int (y-w\sqrt{\gamma})w f^{\rm{No}}(y-w\sqrt{\gamma})f_Z(w)\rd w}{\int f^{\rm{No}}(y-z\sqrt{\gamma})f_Z(z)\rd z}\bigg]\rd x\rd y.
\end{array}\eeqq

Then, taking into account Eqn \eqref{eq:xietabeta},
\beq\label{eq:last} \begin{array}{l}\diy \frac{\rd}{\rd\gamma}h^{\rm w}_{\psi}(Z\sqrt{\gamma}+N)
=\diy \frac{1}{2\sqrt{\gamma}}\diy\int \int \zeta (x,y,\gamma)\xi(y,\gamma) f_{Z,Z\sqrt{\gamma}+N}(x,y)\rd x\rd y\\
\;\qquad -\diy \frac{1}{2\sqrt{\gamma}}\diy \int \int \rho(x,y)\eta(y,\gamma) f_{Z,Z\sqrt{\gamma}+N}(x,y)\rd x\rd y\\
\quad =\diy \frac{1}{2\sqrt{\gamma}}\;\bbE\bigg\{ \zeta_{\rho}(Z, Z\sqrt{\gamma}+N)\xi(Z\sqrt{\gamma}+N,\gamma )-\rho(Z, Z\sqrt{\gamma}+N)\eta(Z\sqrt{\gamma}+N,\gamma )\bigg\}.\end{array}\eeq
$\Box$
\vskip .5 truecm \def\ophi{{\overline\phi}}

{\bf 3.2. WLSI for a WF close to a constant.} Concluding this section, we analyze the WLSI
\eqref{eq:WLSI} when the WF $\phi$ lies in vicinity of a
constant $\ophi$ (and hence, is bounded).
Given independent RVs $X_1$, $X_2$ with PDFs $f_{X_1}$ and $f_{X_2}$, we refer to $Y_1$
and $Y_2$ as $Y_1=X_1/\cos\alpha$, $Y_2=X_2/\sin\alpha$
where $\alpha\in [-\pi,\pi)$ is as in Eqn \eqref{eq:alpha}.
For $Z=Y_1,Y_2$ or  $X_1+X_2=Y_1\cos\alpha+Y_2\sin\alpha$, set:
\beq M(Z;\gamma)={\bbE}\Big[\left\vert Z-{\bbE}\big(Z\big\vert Z\sqrt{\gamma}+N\big)\right\vert^2\Big]
\eeq
and suppose that for the above choices of RV $Z$:
\beq\label{eq:cond4} \bbE \vert \ln f_{Z}(Z)\vert <\infty . 
\eeq
We also need to assume a uniform integrability condition \eqref{eq:cond3}: for an independent
triple $Z,N,N^\prime$ where $N,N^\prime\sim{\rm N}(0,1)$, there exists an
integrable RV $\chi (Z,N)\geq 0$ such that
\beq\label{eq:cond5}\left|\,\log\bbE\left[f_Z\left(Z+\frac{N-N^\prime}{\sqrt\gamma}\right)\Big|Z,N\right]
\,\right|\leq \chi (Z,N).\eeq
\def\wt{\widetilde}

According to formula (4.5) in \cite{KS1}, for these choices of $Z$, the standard SDE $h(Z)$ is
represented as
\beq 
h(Z) 
=h(N)+\frac{1}{2}\int \left[M(Z,\gamma )
-{\bf 1}(\gamma>1)\frac{1}{\gamma}\right]{\rd}\gamma.
\qquad \eeq
Furthermore, as follows from the proof of Theorem 4.1 in \cite{KS1} (see \cite{KS1}, Eqn (4.8)),
for any ${\wt\alpha}\in [-\pi,\pi]$ (including ${\wt\alpha}=\alpha$, the value from  \eqref{eq:alpha}),
\beqq M(Y_1\cos{\wt\alpha}+Y_2\sin{\wt\alpha},\gamma)\geq M(Y_1,\gamma)(\cos\wt\alpha)^2+M(Y_2,\gamma)(\sin\wt\alpha)^2.\qquad \eeqq
For  ${\wt\alpha}=\alpha$, this becomes
\beq\label{eq:refY1Y2}M(X_1+X_2,\gamma)\geq M(Y_1,\gamma)(\cos\alpha)^2+M(Y_2,\gamma)(\sin\alpha)^2.\eeq

Now we are in position to establish Theorem 4 below. As before, we refer to $Z=Y_1,Y_2$ or
$X_1+X_2=Y_1\cos\alpha+Y_2\sin\alpha$.

\vskip.5 truecm

{\bf Theorem 4.} {\sl Let $\gamma_0>0$ be is a point of continuity of
$M(Z,\gamma)$, $Z=Y_1,Y_2,X_1+X_2$. Suppose that there exists
$\delta>0$ such that
\beq\label{eq:Thm4cond} M(X_1+X_2,\gamma_0)\geq M(Y_1,\gamma_0)(\cos\alpha)^2+M(Y_2,\gamma_0)(\sin\alpha)^2+\delta.\eeq
Also assume \eqref{eq:cond4} and \eqref{eq:cond5}.

Then there exists $\eps =\eps (\gamma_0,\delta, f_{X_1}, f_{X_2})$ with the following property.
Let function $x\in\bbR\mapsto\phi(x)\geq 0$ be such that $\left\vert \phi(x)-\ovp\right\vert \leq\eps$,
$\forall$ $x$,
for a  constant $\ovp>0$ . Then the WLSI \eqref{eq:WLSI} with the WF $\phi$ holds true.}
\vskip.5 truecm \def\rC{{\rm C}} \def\rS{{\rm S}}
\noindent

{\bf Proof of Theorem 4.} According to Theorem 1, to prove the WSLI \eqref{eq:WLSI}, we only need to
check that
\beqq (\cos\alpha)^2h^\rw _{\phi_\rC}(Y_1)+(\sin\alpha)^2h^\rw _{\phi_\rS}(Y_2)\leq h^\rw _{\phi}(Y_1
\cos\alpha+Y_2\sin\alpha).\eeqq
For a constant WF $\ovp$, the following inequality is valid (see Ref \cite{KS1}, Lemma 4.2 or Ref \cite{VG},
Eqns (9) and (10))
$$(\cos\alpha)^2h^\rw _{\ovp}(Y_1)+(\sin\alpha)^2h^\rw _{\ovp}(Y_2)\leq h^\rw _{\ovp}(Y_1
\cos\alpha+Y_2\sin\alpha).$$
Next, Theorem 4.1 from \cite{KS1} (applicable because of \eqref{eq:cond4} and \eqref{eq:cond5})
implies that under condition \eqref{eq:Thm4cond}, for
$\eps$ small enough
$$(\cos\alpha)^2h^\rw _{\ovp}(Y_1)+(\sin\alpha)^2h^\rw _{\ovp}(Y_2)+\eps\leq h^\rw _{\ovp}(Y_1
\cos\alpha+Y_2\sin\alpha).$$
Define $\varphi(x)=\vert\phi(x)-\ovp\vert$. It remains to check that
$$h^\rw _{\varphi}(Y_1)\leq \eps/3, h^\rw _{\varphi}(Y_2)\leq \eps/3,  h^\rw _{\varphi}(Y_1
\cos\alpha+Y_2\sin\alpha)\leq \eps/3.$$
But this inequality immediately follows, owing to \eqref{eq:cond4}. This competes the proof of  Theorem 4.
$\quad$ $\Box$
\vskip .5 truecm

The statement of Theorem 4 can be made more efficient for given PDFs $f_{X_1}$ and $f_{X_1}$. As an example, consider the case where RVs $X_1$, $X_2$ are normal and WF $\phi\in C^2$.
\vskip .5 truecm

{\bf Lemma 3.3.} {\sl Let RVs $X_i\sim$N$(\mu_i,\sigma_i^2), i=1,2$ be independent, and $X=X_1+X_2\sim$\\ ${\rm N}(\mu_1+\mu_2,\sigma_1^2+\sigma_2^2)$. Suppose that WF $x\in\bbR\mapsto\phi (x)\geq 0$ is twice contiuously differentiable and slowly
varying in the sense that $\forall$ $x$,
\beq 
\vert\phi''(x)\vert\leq\eps \phi(x),\;\;|\phi(x)-\ophi|<\eps ,\eeq
where $\eps >0$ and $\ophi >0$ are constants.
Then there exists $\eps_0=\eps_0(\mu_0, \mu_1, \sigma_0^2,\sigma_1^2)>0$ such that for any
$0<\eps\leq \eps_0$, the WLSI \eqref{eq01.3} with the WF $\phi$ holds true.}
\vskip 0.5 truecm
\noindent

{\bf  Proof of Lemma 3.3.} Let $\alpha$ be as in Eqn \eqref{eq:alpha};
to  check \eqref{eq01.3},
we use Stein's formula: for $Z\sim{\rm N}(\mu,\sigma^2)$,
\beq\label{eq:stein}{\bbE}\Big[Z^2\phi (Z)\Big]= \sigma^2{\bbE}\Big[\phi(Z)\Big]+ \sigma^4{\bbE}\Big[\phi^{\prime\prime}(Z)\Big].\eeq

Owing to the inequality $|\phi(x)-\ophi|<\eps$,
$$\alpha<\alpha_0=\tan^{-1}\Big(\exp\Big\{(\ophi +\eps)^2[h_+(X_2)-h_-(X_1)]-
(\ophi -\eps)^2[h_+(X_1)-h_-(X_2)]\Big\}\Big).$$
Here
\beqq h_\pm (X_i)=-\bbE\Big[\B1(X_i\in A^i_\pm )\log f^{\rm{No}}_{X_i}(X_i)\Big],\;\; i=1,2,\eeqq
and
\beqq A^i_+=\Big\{x\in\bbR:\; f^{\rm{No}}_i(x)<1\Big\}, \;\;\;A^i_-=\Big\{x\in\bbR:\;f^{\rm{No}}_i(x)>1\Big\}.\eeqq

Evidently,
under conditions $\vert\phi'(x)\vert ,\vert\phi''(x)\vert\leq\eps \phi(x)$ we have that $\alpha_0<\frac{\pi}{2}-\eps$ and $0<\eps< (\sin\alpha)^2, (\cos\alpha)^2<1-\eps<1$. We claim that inequality \eqref{eq01.3}
is satisfied with $\phi$ replaced by $\ophi$ and added $\delta >0$:
\beq\label{eq31.3} \begin{array}{l}
\diy\Big[\log \Big(2\pi (\sigma_1^2+\sigma_2^2)\Big)\Big] \ophi
+\frac{\log e}{\sigma_1^2+\sigma_2^2}\; \ophi\bbE\big[X^2\big]\\
\quad \geq \diy (\cos\alpha )^2\left[\log\frac{2\pi \sigma_1^2}{(\cos\alpha )^2}\right]\ophi +\frac{(\cos\alpha )^2\log e}{\sigma_1^2}\;\ophi\bbE [X_1^2]\\
\qquad + \diy (\sin\alpha )^2\left[\log\frac{2\pi \sigma_2^2}{(\sin\alpha )^2}\right]\ophi +\frac{(\sin\alpha )^2\log e}{\sigma_2^2}\;\ophi\bbE [X_2^2]+\delta.\end{array}\eeq
Here $\delta >0$ is calculated through $\eps$ and increases to a limit $\delta_0>0$ as $\eps\to 0$.

Indeed, strict concavity
of $\log\;y$ for $y>0$ implies that
\beqq \Big[\log \Big(2\pi (\sigma_1^2+\sigma_2^2)\Big)\Big] \geq
(\cos\alpha )^2\left[\log\frac{2\pi \sigma_1^2}{(\cos\alpha )^2}\right]\ophi
+ (\sin\alpha )^2\left[\log\frac{2\pi \sigma_2^2}{(\sin\alpha )^2}\right]+\delta .\eeqq
On the other hand,
\beqq\frac{1}{\sigma_1^2+\sigma_2^2}\; \ophi\bbE\big[X^2\big]=
\frac{(\cos\alpha )^2}{\sigma_1^2}\;\ophi\bbE [X_1^2]
+\frac{(\sin\alpha )^2}{\sigma_2^2}\;\ophi\bbE [X_2^2].
\eeqq

Now, if we look at Eqn \eqref{eq01.3} with WF $\phi$ then, owing to Eqn \eqref{eq31.3} it suffices to
verify that 
\beq\label{eq41.3} \begin{array}{l}
\diy\Big[\log \Big(2\pi (\sigma_1^2+\sigma_2^2)\Big)\Big] \bbE (\phi (X)-\ophi )+\frac{\log e}{\sigma_1^2+\sigma_2^2}\; \bbE\big[X^2(\phi (X)-\ophi )\big]\\
\quad -\diy (\cos\alpha )^2\left[\log\frac{2\pi \sigma_1^2}{(\cos\alpha )^2}\right]\bbE (\phi (X_1)-\ophi ) +\frac{(\cos\alpha )^2\log e}{\sigma_1^2}\;\bbE [X_1^2 (\phi (X_1)-\ophi)]\\
\qquad -\diy (\sin\alpha )^2\left[\log\frac{2\pi \sigma_2^2}{(\sin\alpha )^2}\right]\bbE (\phi (X_2)-\ophi )+\frac{(\sin\alpha )^2\log e}{\sigma_2^2}\;\bbE [X_2^2 (\phi (X_2)-\ophi )]<\delta .\end{array}\eeq
We check this by a brute force, claiming that each term in \eqref{eq41.3} has the absolute value
$<\delta/6$ when $\eps$ is small enough. For the terms containing  $\bbE (\phi (Z)-\ophi )$, $Z=X,X_1,X_2$,
this follows since $|\phi (x)-\ophi |<\eps$.

The terms containing factor $\bbE\big[Z^2(\phi (Z)-\ophi )\big]$
we use Stein`s formula \eqref{eq:stein} and the condition that $|\phi^{\prime\prime}(x)|\leq\eps\phi (x)$.
$\quad$ $\Box$
\vskip .5 truecm

Similar assertions can be established for other examples of PDFs $f_{X_1}$ and $f_{X_2}$.
\vskip 1 truecm

{\emph{Acknowledgements --}}
YS thanks the Math Department, Penn State University,  for the financial support and hospitality
during the academic year 2014-5. SYS thanks the CAPES PNPD-UFSCAR Foundation
for the financial support in the year 2014-5. SYS thanks
the Federal University of Sao Carlos, Department of Statistics, for hospitality during the year 2014-5.
MK acknowledge the support of thew National Research University Higher School of Economics in the framework
of implementation of the 5-100 Pregramme Roadmap.


\begin{thebibliography}{10}

\bibitem{BG} M. Belis and S. Guiasu.  A Quantitative and qualitative measure of information in cybernetic systems. {\it IEEE Trans. on Inf. Theory},  \textbf{14} (1968), 593--594.

\bibitem{C} A. Clim. Weighted entropy with application. {\it Analele Universit\v{a}\c{t}ii Bucure\c{s}ti, Matematic\v{a}}, Anul \textbf{LVII} (2008), 223-231.

\bibitem{CT} T. Cover and J. Thomas. {\it Elements of Information Theory.} New York: Wiley, 2006.


\bibitem{KS1} M. Kelbert and Y. Suhov. Continuity of mutual entropy in the limiting signal-to-noise ratio regimes. In: {\it Stochastic Analysis}, Springer-Verlag: Berlin (2010), 281--299.

\bibitem{KS2} M. Kelbert and Y. Suhov. {\it Information Theory and Coding by Example.} Cambridge: Cambridge University Press, 2013.

\bibitem{L} E.H. Lieb. Proof of entropy conjecture of Wehrl. {\it Commun. Math. Phys.}, {\bf 62} (1978), 35--41.

\bibitem{SY} Y. Suhov and S. Yasaei Sekeh. Simple inequalities for weighted entropies. arXiv 1409.4102.

\bibitem{VG} S. Verd\'{u} and D. Guo. A simple proof of the entropy -power inequality. {\it IEEE Transaction on Information Theory}, \textbf{52}, No. 5 (2006), 2165--2166.




\end{thebibliography}
\end{document}